\documentclass[fleqn,usenatbib]{rasti}

\usepackage{newtxtext,newtxmath}

\usepackage[T1]{fontenc}

\DeclareRobustCommand{\VAN}[3]{#2}
\let\VANthebibliography\thebibliography
\def\thebibliography{\DeclareRobustCommand{\VAN}[3]{##3}\VANthebibliography}

\usepackage{graphicx}	
\usepackage{amsmath}	
\usepackage{xcolor}

\newcommand{\kepler}{\emph{Kepler}}

\newcommand{\plato}{\emph{PLATO}}
\newcommand{\republic}{\texttt{REPUBLIC}}
\newcommand{\matrx}[1]{\ensuremath{\mathrm{#1}}}
\newcommand{\vectr}[1]{\ensuremath{\mathbf{#1}}}
\newcommand{\oscar}[1]{\textcolor{black}{#1}}
\newcommand{\oscartwo}[1]{\textcolor{black}{#1}}

\newcommand{\lbe}{\ensuremath{\lambda_{\rm e}}}
\newcommand{\lbp}{\ensuremath{\lambda_{\rm p}}}
\newcommand{\pgp}{\ensuremath{P_{\rm GP}}}

\title[\texttt{REPUBLIC}]
{\texttt{REPUBLIC}: A variability-preserving systematic-correction algorithm for \plato's multi-camera light curves}

\author[Barrag\'an, Aigrain \& McCormac]{
Oscar~Barragán$^{1}$\thanks{\href{mailto:oscar.barrragan@physics.ox.ac.uk}{oscar.barrragan@physics.ox.ac.uk}},
Suzanne Aigrain$^{1}$,
and James McCormac$^{2,3}$
\\
$^{1}$Sub-department of Astrophysics, Department of Physics, University of Oxford, Oxford, OX1 3RH, UK\\
$^{2}$Physics Department, University of Warwick, Coventry CV4 7AL, UK \\
$^{3}$Centre for Exoplanets and Habitability, University of Warwick, Coventry CV4 7AL, UK
}

\date{Accepted 02/04/2024. Received 27/03/2024; in original form 09/11/2023}

\pubyear{2022}

\begin{document}
\label{firstpage}
\pagerange{\pageref{firstpage}--\pageref{lastpage}}
\maketitle

\begin{abstract}
Space-based photometry missions produce exquisite light curves that contain a wealth of stellar variability on a wide range of timescales. Light curves also typically contain significant instrumental systematics -- spurious, non-astrophysical trends that are common, in varying degrees, to many light curves. Empirical systematics-correction approaches using the information in the light curves themselves have been very successful, but tend to suppress astrophysical signals, particularly on longer timescales. Unlike its predecessors, the \plato\ mission will use multiple cameras to monitor the same stars.
We present \republic, a novel systematics-correction algorithm which exploits this multi-camera configuration to correct systematics that differ between cameras, while preserving the component of each star's signal that is common to all cameras, regardless of timescale.
\oscar{
Through simulations with astrophysical signals (star spots and planetary transits), \kepler-like errors, and white noise, we demonstrate \republic's ability to preserve long-term astrophysical signals usually lost in standard correction techniques. 
We also explore \republic's performance with different number of cameras and systematic properties.}
\oscar{
We conclude that \republic\ should be considered a potential complement to existing strategies for systematic correction in multi-camera surveys, with its utility contingent upon further validation and adaptation to the specific characteristics of the \plato\ mission data.
}
\end{abstract}






\section{Introduction}

The transit method has enabled the detection of thousands of exoplanets so far \cite[see e.g., NASA exoplanet archive\footnote{\url{https://exoplanetarchive.ipac.caltech.edu/}};][]{NASAexoplanet}, and is the most successful exoplanet detection method so far in terms of sheer numbers, thanks in particular to the \emph{Kepler}, \emph{K2} and \emph{TESS} space missions \citep{Borucki2010,Howell2014,Ricker2015}. Typically, transit surveys monitor thousands, if not tens of thousands, of stars simultaneously over a wide field of view. 

\citet{Twicken2010,Stumpe2012,Smith2012} ensure that 
a key component of the data processing chain for these surveys is the removal of common-mode systematic effects. 
These are signals of instrumental (or, for ground-based surveys, telluric) origin that contaminate the light curves of many of the stars to a varying degree, after correction of other known instrumental effects. Methods to identify and correct these effects were first developed in the context of ground-based transit surveys, and typically consist of modelling each light curve as a linear combination of the other light curves observed on the same detector \citep[e.g., Sys-Rem;][]{Tamuz2005}, or of external variables such as seeing and airmass. 
The Sys-Rem method was then refined for \emph{Kepler} to ensure that the extracted systematic trends, usually referred to as Cotrending Basis Vectors (CBVs), are not dominated by the intrinsic variability of the host stars, and to minimize the suppression of these intrinsic variabilities in the corrected light curves \citep{Twicken2010,Stumpe2012,Smith2012}. The resulting methodology, known as "PDC-MAP", represents the current state of the art. It was used for \emph{TESS} with minimal modifications, and forms part of the standard pipeline currently being developed for the \emph{PLATO} (PLAnetary Transits and Oscillations of stars) mission \citep{Rauer2014}.

In contrast to previous space-based transit surveys, \emph{PLATO} uses multiple, small-aperture telescopes with partially overlapping Field Of View (FOV), arranged such that a given star is monitored simultaneously by \oscartwo{6, 12, 18 or 24 cameras, depending on its location in the FOV}. On the ground, the Next Generation Transit Survey \citep[NGTS;][]{wheatley18ngts} also uses multiple telescopes located at the same observatory site, which can be used to monitor the same FOV simultaneously \citep[e.g,][]{Smith2020}. Both \plato\ and NGTS produce per-camera light curves for each star. These are systematics-corrected on a per-camera basis (using variants of the SysRem and PDC-MAP algorithms for NGTS and \plato, respectively) and the light curves from the different cameras are then merged to produce a single light curve with higher signal-to-noise ratio for each star. 

However, as mentioned above, the per-camera systematics removal step tends to suppress some of the intrinsic stellar variability, particularly on timescales approaching the duration of the set of observations being processed. In this paper, we present an alternative systematics-correction method, dubbed \republic, which uses the multiple light curves for a given star across the different cameras, rather than the multiple light curves of all the stars on a given camera. Unlike existing methods, \republic\ explicitly models the stellar signal, which is the same in all the light curves, alongside the systematics, which are specific to each camera. In \republic, all cameras get a `vote' in determining the `true' astrophysical signal. This allows for an efficient separation of any astrophysical signal from the systematics, provided that the latter differ sufficiently from camera to camera.

\section{The \republic\ algorithm}
\label{sec:algo}

\subsection{The model}
\label{sec:model}

Let us assume we have $I$ targets, observed by $J$ cameras, with $K$ observations per camera.
The observed (background-subtracted) flux for target $i$ on camera $j$ in frame $k$ can be written as
\begin{equation}
    F_{ijk} = A_{ik} + B_{ijk} + E_{ijk},
    \label{eq:fijk}
\end{equation}
\noindent
where $A_{ik}$ represents the intrinsic flux of target $i$ in frame $k$, $B_{ijk}$ the systematic noise for target $i$ in frame $k$ on camera $j$, and $E_{ijk}$ the corresponding random noise, which we assume to be white and Gaussian with known standard deviation $\sigma_{ijk}$. We note that $A_{ik}$ does not depend on the camera $j$ as we expect to observe the same astrophysical signal in all cameras. 

Equation \eqref{eq:fijk} assumes that the systematics are additive rather than multiplicative, and the contamination and aperture losses are the same in all cameras. In fact, the relative contribution of nearby contaminant stars to the measured flux for the target does vary from camera to camera. While this is not explicitly accounted for in our formalism, the time-dependent component of the contamination and aperture losses, caused by the drift of the stars' positions over the camera pixel, can be accounted for by the systematics component of the model, which we detail below.

The systematics are modelled as a linear combination of individual `trends', which are specific to each camera but common to all targets on that camera \citep[similar to the PDC-MAP approach;][]{Smith2012}. Specifically, we write the systematic component for each star in each camera as
\begin{equation}
 B_{ijk} = \sum_{n=1}^{N_j} \, W_{ijn} \, T_{jkn},
 \label{eq:bijk}
\end{equation}
\noindent
where $T_{jkn}$ is the value of the $n^{\rm th}$ systematic trend for camera $j$ in frame $k$, $N_j$ is the number of systematic trends per camera $j$, and $W_{ijn}$ is the coefficient linking that trend to target $i$.

\subsection{Trend identification for each camera}
\label{sec:trends}

To implement the model described in the previous section, the first step consists in identifying the systematic trends $T_{jkn}$ for each camera. In this paper, we follow the PDC-MAP approach to extracting CBVs \citep{Stumpe2012,Smith2012}, which consists in selecting the most mutually correlated light curves on a given camera, then using the first few principal components of the resulting matrix as the trends for that camera. The number of CBVs per camera is determined by a user-specified threshold in the fraction of the total variance "explained" by the trends. We note, that one could also use external variables such as airmass and seeing (for ground-based observations), or satellite attitude data, instead of (or in addition to) the CBVs. 

\subsection{Least-squares solution for each star}
\label{sec:solve}

Once the trends are known, it is straightforward to solve for the stellar signal $A_{ik}$ and the systematic trend weights $W_{ijn}$ for a given star $i$ using the fluxes measured for that star on all cameras. \oscartwo{From this point, we drop the subscript $i$ in derivations, since we are considering one star at a time}.

Assuming that the white noise standard deviation is known (which is a reasonable assumption for high-precision photometry), maximising the model likelihood with respect to the parameters is equivalent to minimising the total $\chi^2$.
The total $\chi^2$ for a given target is given by
\begin{equation}
 \chi^2 = \sum_j \sum_k \sigma_{jk}^{-2} \left[F_{jk} - A_{k} - \sum_n W_{jn} T_{jkn} \right]^2.
\end{equation}
Expanding the square:
\begin{equation}
  \label{eq:chi2def}
 \begin{split} 
 \chi^2 = \sum_j \sum_k \sigma_{jk}^{-2} \left[ \vphantom{\sum_{m = 1}^{N_j} } \right. &  
 F^2_{jk} + A_{k}^2 - 2 F_{jk} A_{k}  \\
 & + \sum_{n=1}^{N_j} W_{jn} T_{jkn} \sum_{m = 1}^N W_{jm} T_{jkm} \\ 
 & \left. + 2 \left( A_{k} - F_{jk} \right) \sum_{n=1}^{N_j} W_{jn} T_{jkn} \right]. 
 \end{split}
\end{equation}
To solve for the parameters, we evaluate the derivatives of Equation~\eqref{eq:chi2def} with respect to $A_{k}$ and $W_{jn}$ and set them to zero:
\begin{equation}
\label{eq:dchi2da}
   0 = \frac{\partial \chi^2}{\partial A_{k}} = 2 A_{k} \sum_j \frac{1}{\sigma_{jk}^2} - 2 \sum_j \frac{F_{jk}}{\sigma_{jk}^2} + 2 \sum_j \sum_n \frac{W_{jn} T_{jkn}}{\sigma_{jk}^2}
\end{equation}
and
\begin{equation}
\label{eq:dchi2dw}
\begin{split}
0 = \frac{\partial \chi^2}{\partial W_{jn}} & = &
2 W_{jn} \sum_k \frac{ T_{jkn}^2}{\sigma_{jk}^2} + 2 \sum_{m \ne n} W_{jm} \sum_k \frac{T_{jkn} T_{jkm}}{\sigma_{jk}^2} \\
& & + 2 \sum_k \frac{A_{k} T_{jkn}}{\sigma_{jk}^2} - 2 \sum_k \frac{F_{jk} T_{jkn}}{\sigma_{jk}^2}.
\end{split}
\end{equation}


\subsection{Matrix formulation}
\label{sec:matrix}

We now re-arrange Equations~\eqref{eq:dchi2da} and \eqref{eq:dchi2dw}, and combine them into a single matrix equation of the form 
\begin{equation}
    \matrx{X} \, \vectr{p} = \vectr{y},
\end{equation}
where $\vectr{p}$ is a vector containing all the unknowns quantities, i.e.\ the free parameters of the model, namely the $A_k$'s and $W_{jn}$'s, while the matrix $\matrx{X}$ and the vector $\vectr{y}$ depend only on the known values $F_{jk}$, $\sigma_{jk}$ and the $B_{jkn}$. Specifically, the vector $\vectr{p}$ is given by
\begin{equation}
    \vectr{p} = \left(\vectr{a}, \vectr{w}_1, \vectr{w}_2, \ldots, \vectr{w}_j, \ldots, \vectr{w}_J \right)^{\rm T},    
\end{equation}
where 
\begin{equation}
    \vectr{a} = \left(A_1, A_2, \ldots, A_k, \ldots, A_K \right)^{\rm T}
\end{equation}
and 
\begin{equation}
    \vectr{w}_J = \left(W_{j1}, W_{j2}, \ldots, W_{jn}, \ldots, W_{jN_j} \right)^{\rm T}.
\end{equation}
The matrix $\matrx{X}$ is given by
\oscartwo{
\begin{equation}
 \label{eq:bigmatrix} 
 \matrx{X}=
 \begin{bmatrix} \matrx{C} & \matrx{D}_1 & \matrx{D}_2 & \dots & \matrx{D}_J \\[1mm]
    \matrx{D}_1^T & \matrx{E}_1 & 0 & \dots & 0 \\[1mm]
    \matrx{D}_2^T & 0 & \matrx{E}_{2} & \dots & 0 \\[1mm]
    \vdots & \vdots & \vdots & \ddots & \vdots \\[1mm]
    \matrx{D}_J^T & 0 & 0 & \dots & \matrx{E}_J 
  \end{bmatrix},
\end{equation}
}
where
\begin{equation}
    \matrx{C} = {\rm diag} \left\{ \sum_j \sigma_{jk}^{-2},~k=1 \rightarrow K\right\},
    \label{eq:bigc}
\end{equation}
\begin{equation}
  \matrx{D}_j = \begin{bmatrix} 
    \frac{T_{j11}}{\sigma_{j1}^2} &  \frac{T_{j12}}{\sigma_{j1}^2} &  \dots &  \frac{T_{j1N_j}}{\sigma_{j1}^2} \\ 
    \frac{T_{j21}}{\sigma_{j2}^2} &  \frac{T_{j22}}{\sigma_{j2}^2} &  \dots &  \frac{T_{j2N_j}}{\sigma_{j2}^2} \\ \vdots &  \vdots &  \ddots &  \vdots \\ \frac{T_{jK1}}{\sigma_{jK}^2} &  \frac{T_{jK2}}{\sigma_{jK}^2} &  \dots &  \frac{T_{jKN_j}}{\sigma_{jK}^2} 
    \label{eq:bigd}
  \end{bmatrix},    
\end{equation}
and
\begin{equation}
  \matrx{E}_j = \begin{bmatrix} 
    \sum_k \frac{T_{jk1}^2}{\sigma_{jk}^2} &  \sum_k \frac{T_{jk1} T_{jk2}}{\sigma_{jk}^2} &  \dots &  \sum_k \frac{T_{jk1} T_{jkN_j}}{\sigma_{jk}^2} \\ 
    \sum_k \frac{T_{jk1} T_{jk2}}{\sigma_{jk}^2} &  \sum_k \frac{T_{jk2}^2}{\sigma_{jk}^2} &  \dots &  \sum_k \frac{T_{jk2} T_{jkN_j}}{\sigma_{jk}^2} \\ 
    \vdots &  \vdots &  \ddots &  \vdots \\ 
    \sum_k \frac{T_{jk1} T_{jkN_j}}{\sigma_{jk}^2} &  \sum_k \frac{T_{jk2} T_{jkN_j}}{\sigma_{jk}^2} &  \dots &  \sum_k \frac{T_{jkN_j}^2}{\sigma_{jk}^2} 
  \end{bmatrix}.
   \label{eq:bige}
\end{equation}
Finally, the vector $\vectr{y}$ is given by 
\begin{equation}
\vectr{y} = \left(\vectr{q}, \vectr{r}_1, \vectr{r}_2, \ldots, \vectr{r}_j, \ldots, \vectr{r}_J \right)^{\rm T},    
\end{equation}
where
\begin{equation}
\vectr{q} = \left( \sum_j \frac{F_{j1}}{\sigma_{j1}^2}, \sum_j \frac{F_{j2}}{\sigma_{j2}^2} \\ \ldots, \sum_j \frac{F_{jK}}{\sigma_{jK}^2} \right)^{\rm T}    
\end{equation}
is the variance-weighted average of the observed flux across all cameras,
and
\begin{equation}
 \vectr{r}_j = \left( \sum_k \frac{F_{jk} T_{jk1}}{\sigma_{jk}^2}, \sum_k \frac{F_{jk} T_{jk2}}{\sigma_{jk}^2},  \ldots, \sum_k \frac{F_{jk} T_{jkN}}{\sigma_{jk}^2} \\ \right)^{\rm T}.    
\end{equation}

\subsection{Inverting the matrix \matrx{X}}
\label{sec:block}

\oscartwo{We can now rewrite eq.~\eqref{eq:bigmatrix} in order to simplify the solution of the linear problem. $\matrx{X}$ can be rewritten as a $2\times2$ block matrix of the form}

\begin{equation}
    \matrx{X} =\begin{bmatrix} \matrx{C} &  \matrx{D} \\ \matrx{D}^T &  \matrx{E}\end{bmatrix},
\end{equation}
\oscartwo{where $C$, $D$, and $E$ are computed from equations \eqref{eq:bigc},\eqref{eq:bigd} and \eqref{eq:bige}, respectively. The inverse of $X$ can then ben computed as} \citep[see e.g.,][]{Lu2002}
\begin{equation}
 \matrx{X}^{-1} = \begin{bmatrix} \matrx{C}^{-1} + \matrx{C}^{-1} \, \matrx{D} \, \matrx{S}^{-1}\,  \matrx{D}^T \, \matrx{C}^{-1} &  - \matrx{C}^{-1} \, \matrx{D} \, \matrx{S}^{-1} \\
 - \matrx{S}^{-1} \, \matrx{D}^T \, \matrx{C}^{-1} &  \matrx{S}^{-1} \end{bmatrix},
 \label{eq:xinverse}
\end{equation}
where
\begin{equation}
    \matrx{S} = \matrx{E} -\matrx{D}^T \, \matrx{C}^{-1} \, \matrx{D},
\end{equation}
is the Schur complement of the block $\matrx{C}$. Thus, to evaluate $\matrx{X}^{-1}$ we only need to invert the matrices $\matrx{C}$ and $\matrx{S}$.
As $\matrx{C}$ is diagonal, its inverse is given by
\begin{equation}
     \left(\matrx{C}^{-1}\right)_{k} = \left(\matrx{C}_{k}\right)^{-1}= \left( \sum_j \sigma_{jk}^{-2} \right)^{-1}.
\end{equation}
Furthermore, $\matrx{S}$ is Hermitian and positive-definite, so its inverse can be computed efficiently using Cholesky decomposition. \oscar{We note that the size of matrix $\matrx{E}$ is independent of the number of observations, while the size of matrix $\matrx{D}$ scales quadratically with the number of observations, but it never needs to be inverted. Therefore, 
the computational cost of \republic\ scales linearly with the number of observations, and cubically with the total number of systematic trends across all cameras, while its memory usage scales cubically with the number of observations, and linear with the number of systematic trends.}

We now have all the ingredients to compute
\begin{equation}
    \label{eq:solve}
    \vectr{p} = \matrx{X}^{-1} \, \vectr{y},
\end{equation}
and thus extract simultaneously the best-fit astrophysical signal $\vectr{a}$ and the weights of the systematics model in each camera, $\vectr{w}_j$. We can also construct a systematics-corrected light curve in each camera,
\begin{equation}
    \vectr{\tilde{f}}_j = \vectr{f} - \vectr{w}^T_j \, \matrx{T}_j.
\end{equation}

\section{Tests}

As a proof of concept, we created multi-camera, \plato-like synthetic light curves containing astrophysical signals, systematic trends, and white noise. We then applied the algorithm described in section~\ref{sec:algo} and compared the results to a naive (least-squares) implementation of the standard PDC algorithm, which corrects the light curves from each camera separately.

\oscar{There two main tools currently available to simulate large numbers of \plato\ light curves. PlatoSim \citep{platosim} is a full, end-to-end simulator, which starts by generating full images including astrophysical signals and all known instrumental effects, then mimics the pixel-level correction and light curve extraction and pre-processing implemented in the \plato\ level 0 and level 1 pipelines. However, at the time of writing, the behaviour of the instrument remains limited, only a few of the cameras have been fully tested, nd the level 1 pipeline is not yet fully implemented in PlatoSim. The only multi-camera, multi-quarter dataset produced with PlatoSim to far \citep{platoSim3} was generated using pessimistic assumptions concerning some of the key instrumental noise sources, resulting in strong systematics that are common to all cameras, which makes it unsuitable to test \republic. Alternatively, the \plato\ Solar-like Light curve Simulator \citep[PSLS;][]{psls} is a lighter-weight tool that uses semi-analytic approximations to generate systematics and noise time-series to which the stellar signals are added separately. It produces outputs that are entirely uncorrelated between cameras, which would constitute an un-realistically favourable test for \republic. Therefore, we opted to produce our own, simpler simulations, where we control every aspect of the process, but note that \republic\ should be tested on more realistic simulated data once the behaviour of the satellite  and instrument are better understood, and the pipelines have been finalised.}

\subsection{Light curve generation}
\label{sec:lcs}

\subsubsection{Astrophysical signals}

We used \texttt{citlalicue} \citep[][]{pyaneti2} to create synthetic stellar light curves containing transit signals and stellar activity signals. Granulation and stellar oscillations were not included in our simulations, as we wished to focus on astrophysical signals with timescales similar to both the dominant systematics and the duration of the simulated observations, but we note that the \republic\ model is entirely agnostic regarding the nature and timescales of the astrophysical signal.

Each light curve consists of 1000 equally spaced observations over 90 days (corresponding to one \plato\ quarter). Note that this results in an interval between consecutive observations of just over 2h, significantly longer than the cadence of real \plato\ observations (which will vary between 2.5s and 600s, depending on the stellar sample). This longer cadence is sufficient for a proof-of-concept, and results in much smaller data volumes. However, we note that \republic\ is agnostic of the sampling of the data, its performance will be the same even in the 2.5s cadence. \oscar{While we do not report on these tests here for the reasons outlined above, we have run \republic\ on up to 8 quarters of simulated \plato\ data at 25s cadence, without encountering any CPU or memory usage problems.} 

To mimic the stellar activity signals, we use random samples from a Gaussian Process (GP) with a Quasi-Periodic kernel:
\begin{equation}
    K(t_i,t_j) = A^2 \exp 
    \left[
    - \frac{(t_i - t_j)^2}{2\lambda_{\rm e}^2}
    - \frac{\sin^2[\pi(t_i - t_j)/P_{\rm GP}]}{2 \lambda_{\rm P}^2}
    \right]
    \label{eq:gamma_normal}
\end{equation}
where $A$ is the amplitude (or standard deviation) of the GP; \pgp\ is its characteristic period; \lbp\ is the characteristic length scale of the periodic component (the harmonic complexity $\Gamma$ is related to \lbp\ via $\Gamma = 1/2\lbp^2$); and \lbe\ is the characteristic evolution timescale. For each light curve, these hyper-parameters were drawn randomly from uniform distributions in the following intervals:
 $A\in[1\times10^{-5},5\times10^{-3}]$\,ppm; $\lbe\in[10,1000]$\,days; $\lbp\in[0.1,2]$; and $\pgp\in[3,30]$\,days. 

We also include one planetary transit signal per light curve. The parameters of the injected planetary signals were also drawn from uniform distributions with the following intervals: transit epoch $T_0\in[0,5]$\,days; orbital period $P\in[0,10]$\,days; impact parameter $b\in[0,1]$; scaled semi-major axis, $a/R_\star\in[1.5,50]$; scaled planet radius $r_{\rm p}/R_\star\in[0.01,0.1]$. We assumed all orbits to be circular, and did not include limb-darkening. An example of one of the synthetic stellar signals is shown in panel a) of Figure~\ref{fig:republic}. 


\subsubsection{Systematic trends and white noise}
\label{sec:sim_trends}

We then produced multiple copies of each stellar signal, to which we added different simulated systematic effects to represent the observations taken by the different cameras. In the case of \plato, the 24 cameras are arranged in 4 groups of 6 cameras each. Each camera group observes the same FOV and timestamps, with a new integration starting every 25s. On the other hand, the FOV of the different camera groups overlap only partially, and their timestamps are staggered relative to each other (by 6.25s) to allow for more efficient use of the readout electronics and processing power. In this paper, we ignore the fact that the timestamps are not simultaneous, but we note that the light curves from different camera groups would need to be interpolated onto the same grid of time-stamps before applying \republic. 

\begin{figure*}
    \centering
    \includegraphics[width=0.98\textwidth]{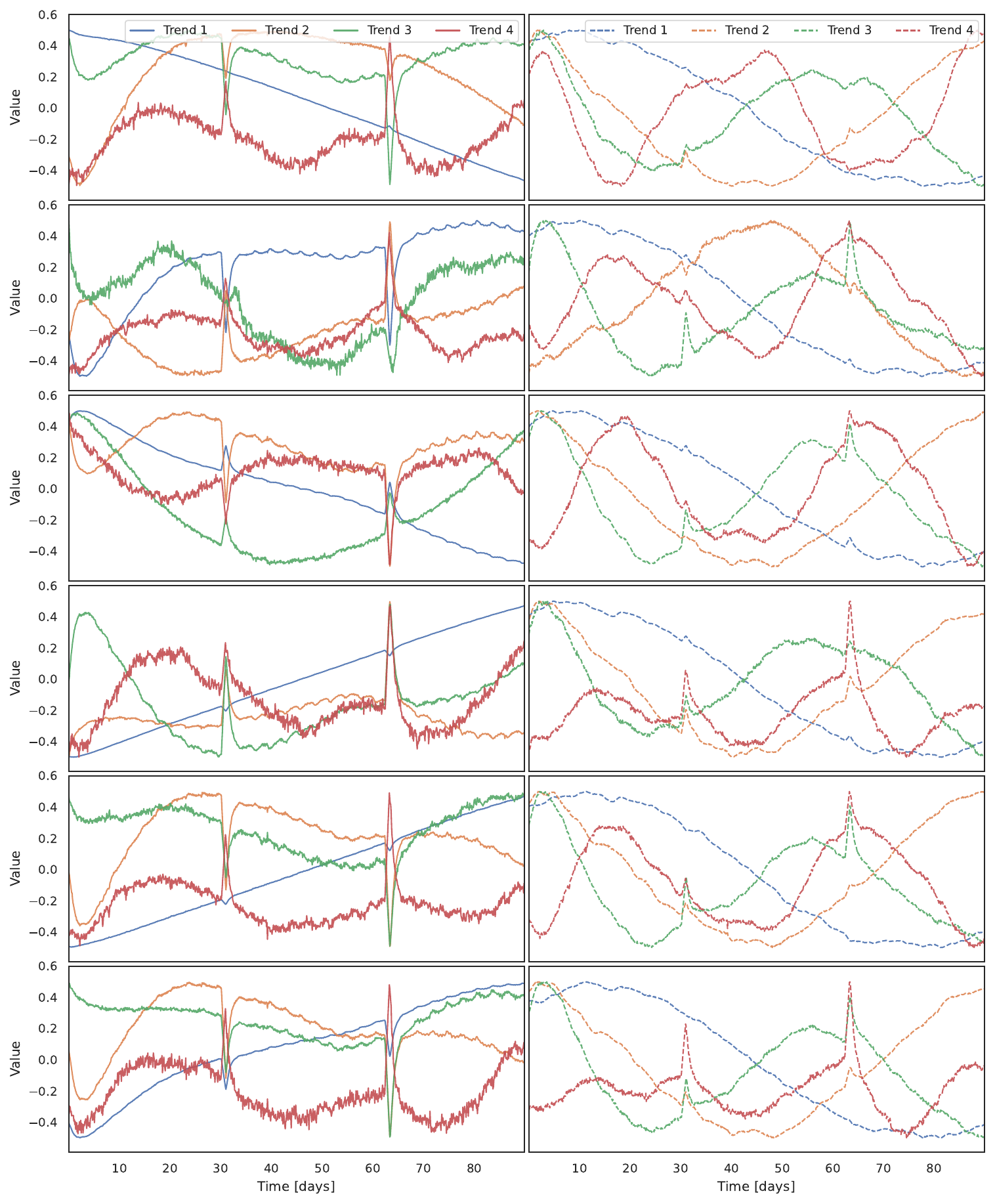}
    \caption{\emph{Left panel:} Systematic trends added to our simulated light curves. These trends correspond to \kepler\ CBVs from its 10th quarter.
    Each panel (from top to bottom) corresponds to a different camera, while each colour represents a different trend. 
    \emph{Right panel:} Recovered systematic trends using PCA decomposition. Panels and colours represent the same cameras and trends as in the left panel.
    }
    \label{fig:trends}
\end{figure*}

To make our simulated systematics trends as realistic as possible, we used the \texttt{lightkurve} package \citep[][]{lightkurve} to download CBVs extracted from the \kepler\ data by the PDC-MAP pipeline. We used the CBVs obtained for the tenth quarter of the mission, but we emphasise that this selection is arbitrary and these tests work with any other \kepler\ quarter. For each camera, we selected one of the 80 \kepler\ readout channels at random and used the first 4 CBVs from that channel as our systematic trends. We then assigned a random weight, between $-50$ and $50$\,ppt, to each CBV in each camera. 
In order to account for the difference in time-sampling between the \kepler\ data and our simulations, we interpolated the original CBVs to the time stamps of the synthetic light curves.
In Figure~\ref{fig:trends}, we present an illustration of the four distinct systematics observed in the \kepler\ dataset as they are included in six of our simulated \plato-like light curves. 
Notably, these signals exhibit discernible trends and manifest abrupt discontinuities approximately every 30 days. These discontinuities correspond to the interruptions in \kepler\ observations to enable data to be transmitted back to Earth. 

\begin{figure*}
    \centering
    \includegraphics[width=0.98\textwidth]{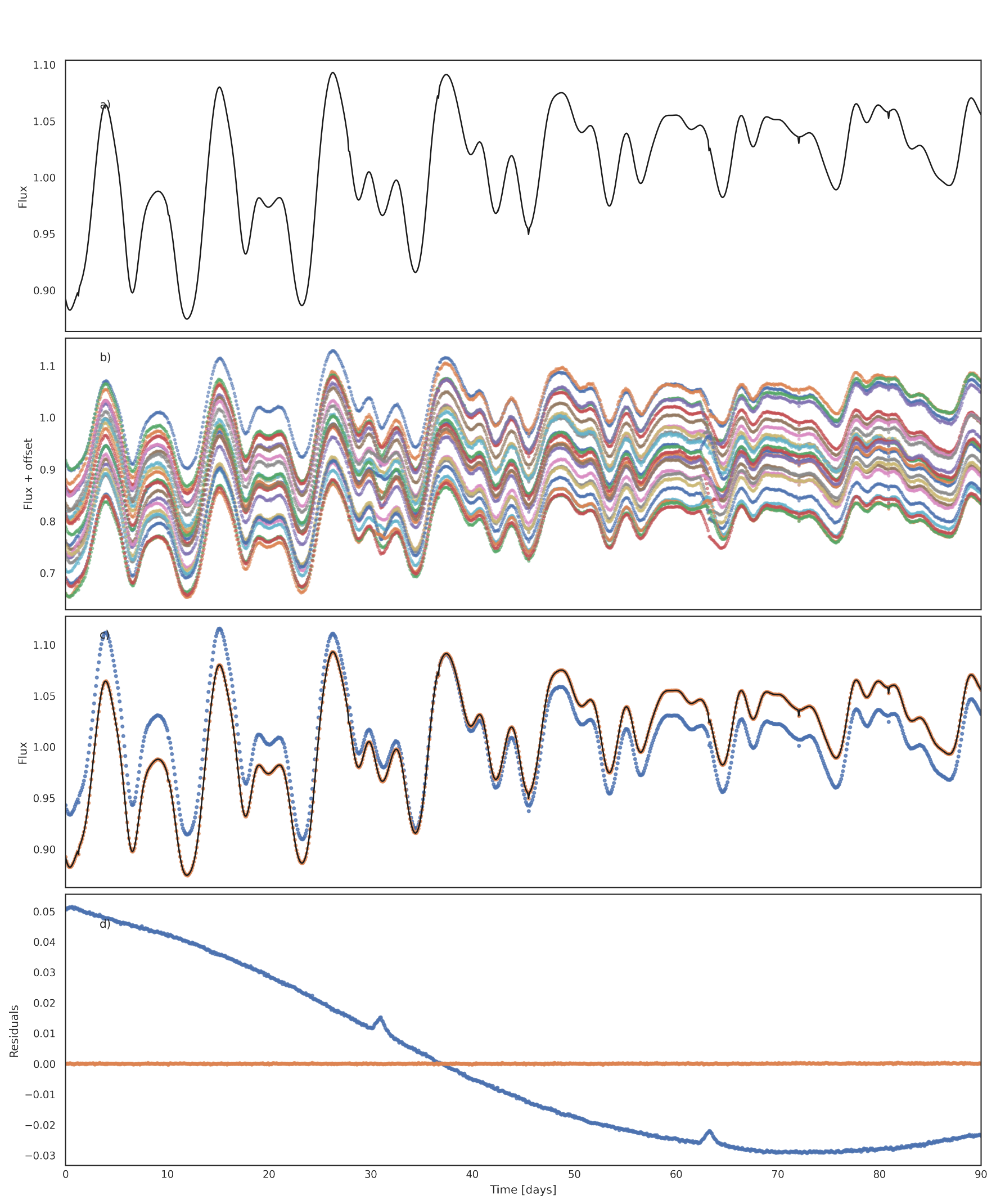}
    \caption{Example application of \protect\republic\ to a simulated multi-camera light curve. Panel a) shows the synthetic astrophysical signal. Panel b) shows the light curves for the 24 simulated cameras. They contain systematics and white noise as described in the text, in addition to the astrophysical signal and an offset to help visibility. Panel c) shows the recovered astrophysical signal using the \republic\ (orange circles) and PDC-LS-like (blue circles) algorithms. We also show the true signal (solid black line) for comparison. Panel d) shows the difference between the merged, systematics-corrected light curve and the injected astrophysical signal, using \republic\ (orange circles) and PDC-LS (blue circles).
    }
    \label{fig:republic}
\end{figure*}

Finally, we added Gaussian white noise to the data, maintaining a standard deviation of 500 ppm for each individual exposure across all cameras. Panel b) of Fig.\ref{fig:republic} depicts a set of 24 synthetic multi-camera light curves corresponding to a single target, including the aforementioned white noise.

\subsection{Results in the fiducial case}
\label{sec:idealcase}

Our first set of simulations used different CBVs as the systematic trends in each camera, so that the resulting systematics were not strongly mutually correlated. Furthermore, we initially assumed that these trends were known perfectly, by setting the $T_{jkn}$ in the \republic\ model equal to the CBVs used to generate the light curves. We then fit for the systematic weights $W_{ijn}$ alongside the astrophysical signal $A_{ik}$ as described in Sections~\ref{sec:solve}--\ref{sec:block}. For comparison, we also perform a least-squares fit for the systematic weights in each camera separately, ignoring the astrophysical signal, before averaging the resulting per-camera light curves. Hereafter, we refer to this approach as PDC-LS, after the least-squares version of the PDC algorithm, which it approximates. 

Panel c) of Figure~\ref{fig:republic} shows an example of the signals recovered by both algorithms. In general, both algorithms result in per-camera light curves with typical point-to-point scatter (measured from the first difference of the light curve, which preserves only short-term signals) of $\sim 500 $\,ppm, which averages down to \oscar{$\sim 115$\,ppm,} for 24 cameras. This is the expected behaviour (following $\approx \sigma/\sqrt{J}$, where $J$ is the number of cameras) and it shows that both algorithms treat the white noise component of the signal in the same way.

Short-term scatter aside, the injected astrophysical signals, such as star-spot modulation and/or transits, are visible in the corrected light curves produced by both methods, but the long-term component of the stellar signal is suppressed in the PDC-LS version, whereas it is preserved in the \republic\ version. To emphasise this, panel d) of Figure~\ref{fig:republic} shows the difference between the corrected, merged light curves and the true astrophysical signal. Only white noise is present in the \republic\ version, whereas the PDC-LS-like method has `injected' a spurious long-term signal into the corrected light curve. 
\oscar{It is also evident that the PDC-LS method exhibits minor fluctuations at $30$ and $60$ days, attributable to the \kepler\ trends incorporated into the light curves. These perturbations are clearly observed in Figure~\ref{fig:trends} (they will also appear in the residuals of other figures in the paper).
}
We only show an example in this paper, but we note that this behaviour was seen consistently across the entire set of 1000 simulated stars.

\subsection{Sensitivity to number of cameras and white noise}

The next step was to test how \republic\ performs with a smaller number of cameras. Besides the 24 cameras case described before, we also tested for 6 and 12 detectors. While stars located near the centre of the \plato\ Field-Of-View (FOV) will be observed by all 24 cameras, other parts of the FOV will be observed by only 18, 12, or 6 cameras. We found that the nominal white noise of 500\,ppm averages down to \oscar{$\sim 220$\,ppm and $\sim 150$\,ppm} for 6 and 12 cameras, respectively. This behaviour is as expected, similar to the case with 24 cameras. 

\begin{figure*}
    \includegraphics[width=0.98\textwidth]{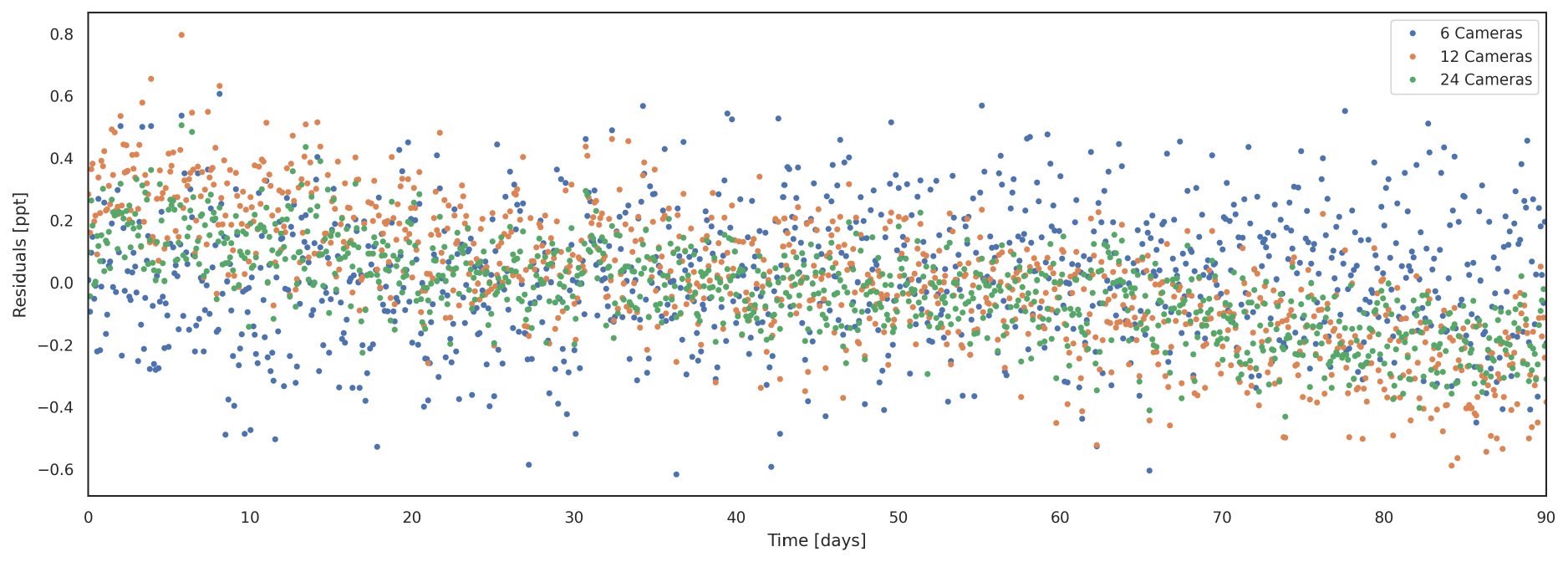}\\
    \includegraphics[width=0.98\textwidth]{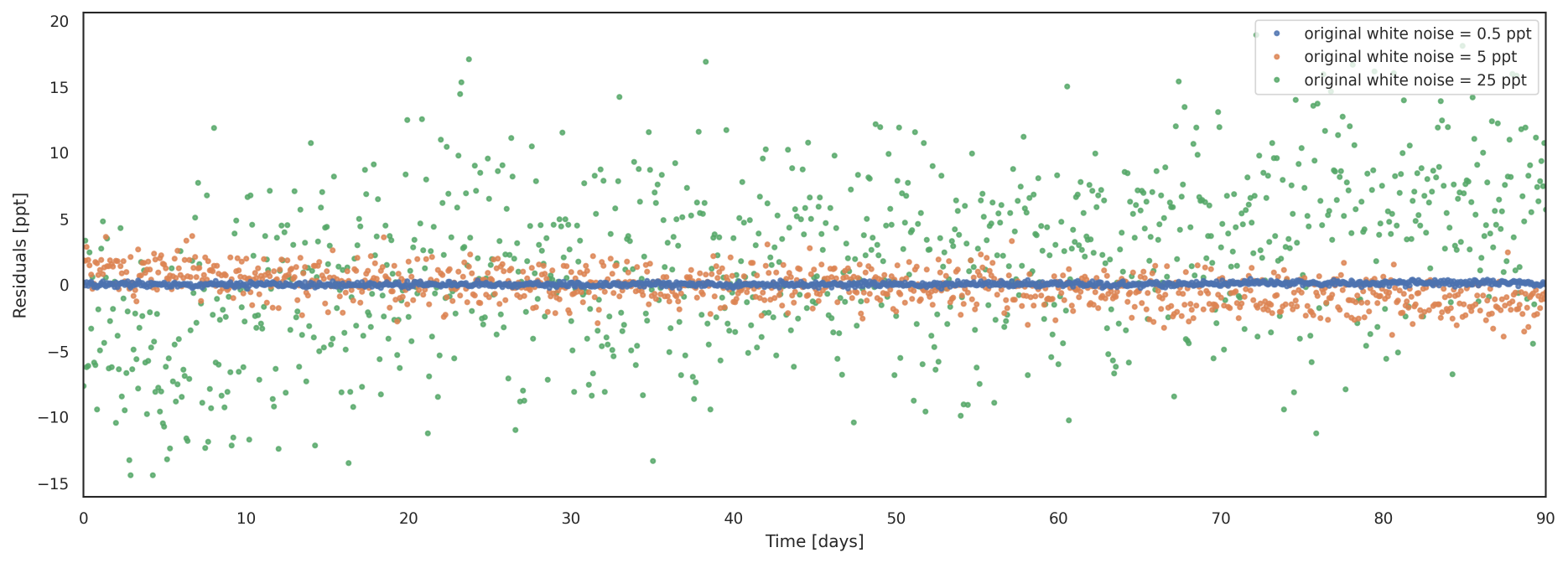}
    \caption{Dependence of the performance on the number of cameras (top) and the white noise level per camera (bottom). In both panels, we show the difference between the \republic-corrected, merged light curve and the true astrophysical signal, leaving behind white noise and residual systematic trends. The white noise level was fixed at 0.5 ppt in the top panel, and the number of cameras at 24 in the bottom one. 
    Note that this example corresponds to the same light curve presented in Fig.~\ref{fig:republic}.
    }
    \label{fig:rescameras}
\end{figure*}

The top panel of Figure~\ref{fig:rescameras} shows the residuals (\republic-corrected light curve minus true astrophysical signal) for the 6, 12, and 24 camera cases for a randomly selected example light curve. We can see visually the improvement in white noise when increasing the number of cameras. In some cases, we also observed that residual systematic trends start to appear in the \republic-corrected light curves as the number of cameras decreases, although this is partly masked by the larger white noise.  This behaviour is entirely in line with expectations.

We also explore the behaviour of the \republic\ correction when light curves contain higher levels of white noise. 
We reprocessed the light curves created in Section~\ref{sec:lcs} to add white noise of 5 and 25 ppt (significantly larger than the nominal 0.5 ppt that we used before). We applied a \republic\ correction to each set of light curves and we found that the noise goes down to 1.1 ppt and 5.8 ppt for the light curves with 5 ppt and 25 ppt noise levels, respectively. This behaviour is nominal, scaling approximately as $\sigma/\sqrt{J}$, where $J = 24$.
An example of this test is shown in the bottom panel of Figure~\ref{fig:rescameras}. Despite the increase in residual dispersion, increasing the white noise level does not affect \republic's ability to correct the systematic trends (provided, of course, the trends themselves are known -- we address that point in the next section). 
\newline

\oscartwo{We have shown then that in the ideal scenario where the systematics across all cameras are distinct, the \republic\ algorithm demonstrates robust performance, unaffected by either the number of cameras or variations in white noise levels. 
Consequently, even with a reduced number of cameras or in the presence of increased white noise, \republic's efficacy in systematic correction remains consistently high, showcasing its potential for application in diverse observational conditions.}

\subsection{Breaking assumptions}

Until this point, we have tested the \republic\ in idealised simulations, demonstrating that the algorithm functions as expected. However, several of the assumptions we have made cannot be guaranteed in practice. In particular, we assumed that the systematic trends were perfectly known and mutually uncorrelated. In a real-world scenario, the dominant trends are unknown and are typically extracted from the ensemble of light curves itself \citep[see e.g.,][]{Smith2012}. Furthermore, at least some sources of error affect the different cameras in the same way, leading to systematics that can be strongly correlated across the different cameras. In this subsection, we explore how these real-world constraints affect the performance of the \republic\ algorithm.

\subsubsection{Unknown trends}
\label{sec:pcatrends}


Instead of assuming the trends were known \emph{a-priori}, we attempted to extract them from the sample of 1000 light curves in 24 cameras created in Section~\ref{sec:lcs}. For each camera, we perform a Principal Component Analysis (PCA) decomposition of the set of 1000 LCs, and extract the first 4 Principal Components to be used as estimated trends. 
Figure~\ref{fig:trends} shows the recovered trends for six of the simulated cameras. As we can see, the PCA-inferred trends provide a good, but imperfect approximation to the true (\kepler\ CBV-derived) trends.

\begin{figure*}
    \includegraphics[width=0.98\textwidth]{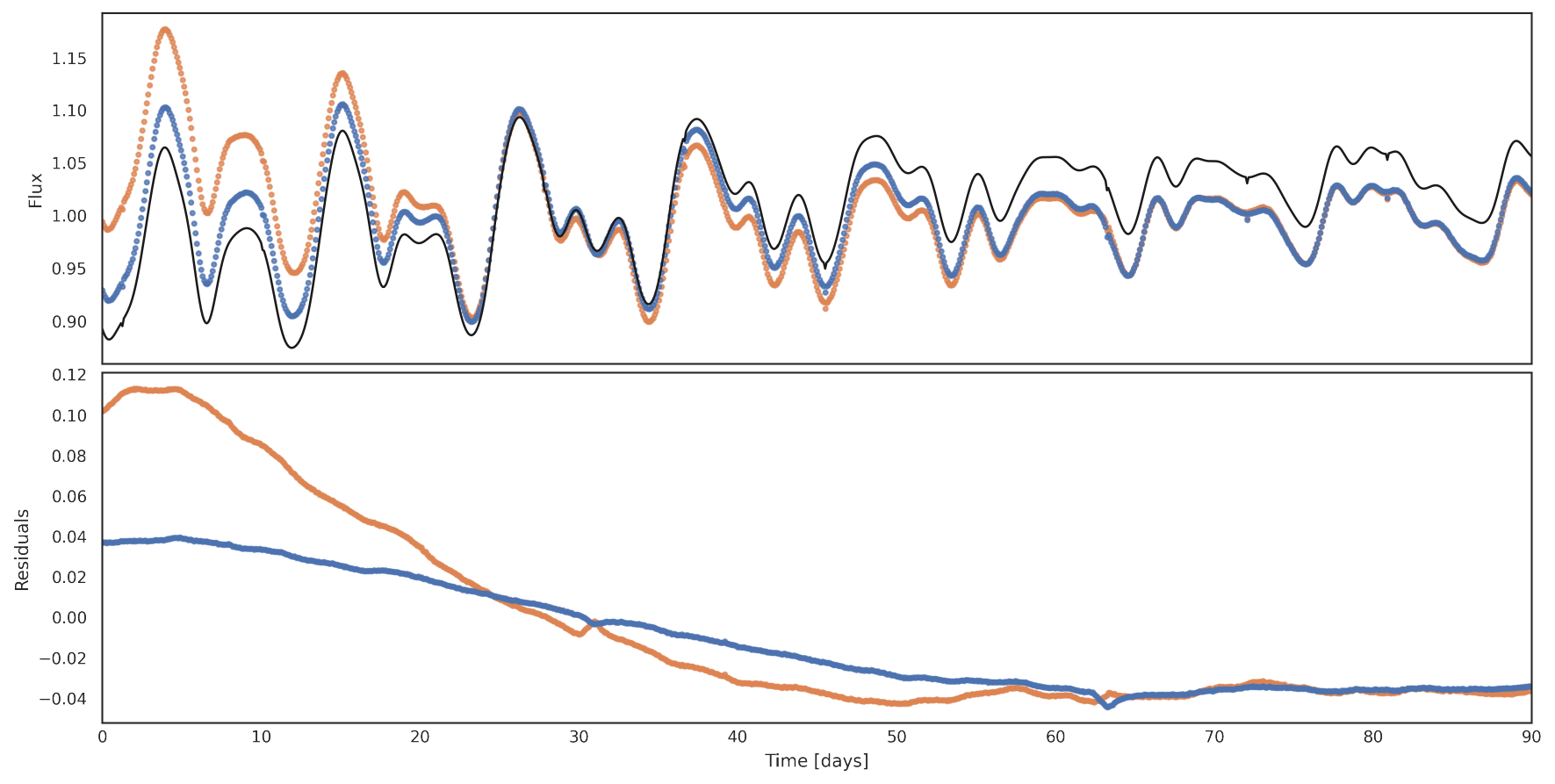}
   \caption{Impact of \oscar{imperfectly-known trends} on \protect\republic's performance. The top panel shows the merged, \oscar{\republic}-corrected light curve with the PCA-extracted trends in orange, compared to the true astrophysical signal in black. 
   We also show the performance of the PDC-LS-like algorithm using these imperfect trends in blue.
   The bottom panel shows the corresponding residuals for each case. 
    }
    \label{fig:lcnonideal}
\end{figure*}

We then used these PCA-inferred trends as the $T_{jkn}$ in the \republic\ algorithm to correct the light curves. 
Figure~\ref{fig:lcnonideal} shows the recovered astrophysical signal using the PCA-inferred trends for a given light curve, compared to the ideal case where the trends are known. The recovered astrophysical signal retains the bulk of the intrinsic stellar and planetary signals, but the imperfect knowledge of the underlying systematic trends results in significant residual long-term trends that are not of astrophysical origin. 
We also performed a PDC-LS approach to correct the light curves using the PCA-inferred trends. Figure~\ref{fig:lcnonideal} shows the recovered astrophysical signal in this case. We can see that, as in the \republic\ case, the imperfect trends leave a residual signal in the recovered astrophysical signal. 
These results highlight the importance of the trend extraction in the systematics-correction process. 

It is worth noting that the light curve sample used in this test is not entirely realistic. For example, all the light curves share the same white noise level and the same parent parameter distributions for the stellar and planetary signals. Furthermore, the method employed here to extract the trends is very simplistic compared to the procedure used in the PDC-MAP pipeline \citep{Stumpe2012}, which is also foreseen to be used for \plato. The PDC-MAP trend extraction also uses PCA, but works with a sub-set of the light curves pre-selected to be mutually correlated. This pre-selection ensures that the extracted CBVs are not excessively affected by the variability of individual stars. In our sample, all the simulated light curves contain planetary signals and significant stellar variability, making the identification of a "low variability" or systematic-dominated subset impossible. A more sophisticated trend extraction approach applied to a more realistic light curve sample would likely result in a better set of estimated trends, so the results in this section should be seen as a worst-case scenario.


\subsubsection{\oscar{Similar} trends across multiple cameras}

\begin{figure*}
    \centering
    \includegraphics[width=0.98\textwidth]{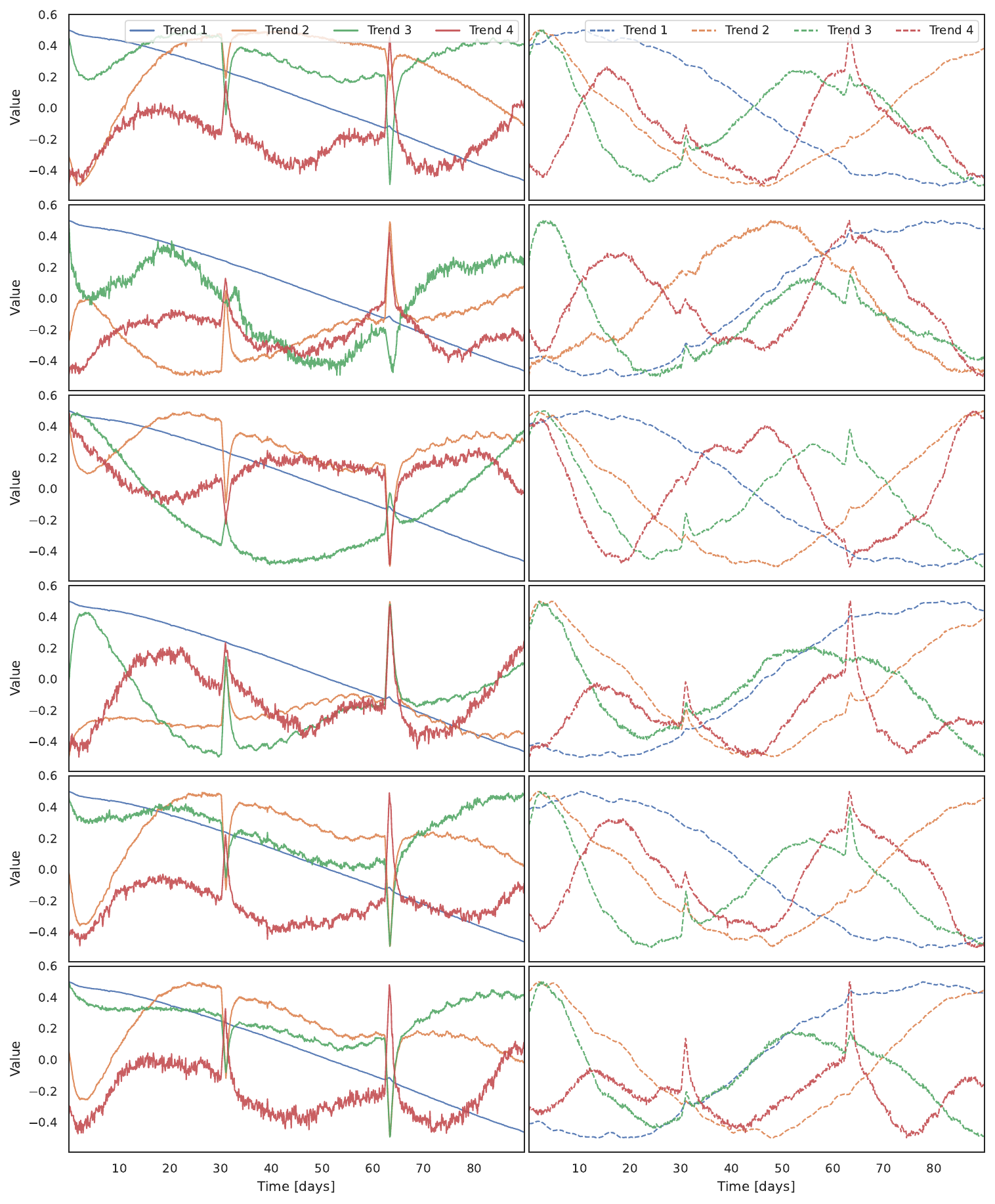}
    \caption{
    Plots are the same as in Figure~\ref{fig:trends}, for the identical trend case.
    In the left panel, one of the trends (blue) is identical for all the cameras.
    }
    \label{fig:trends_correlated}
\end{figure*}

\begin{figure*}
    \includegraphics[width=0.98\textwidth]{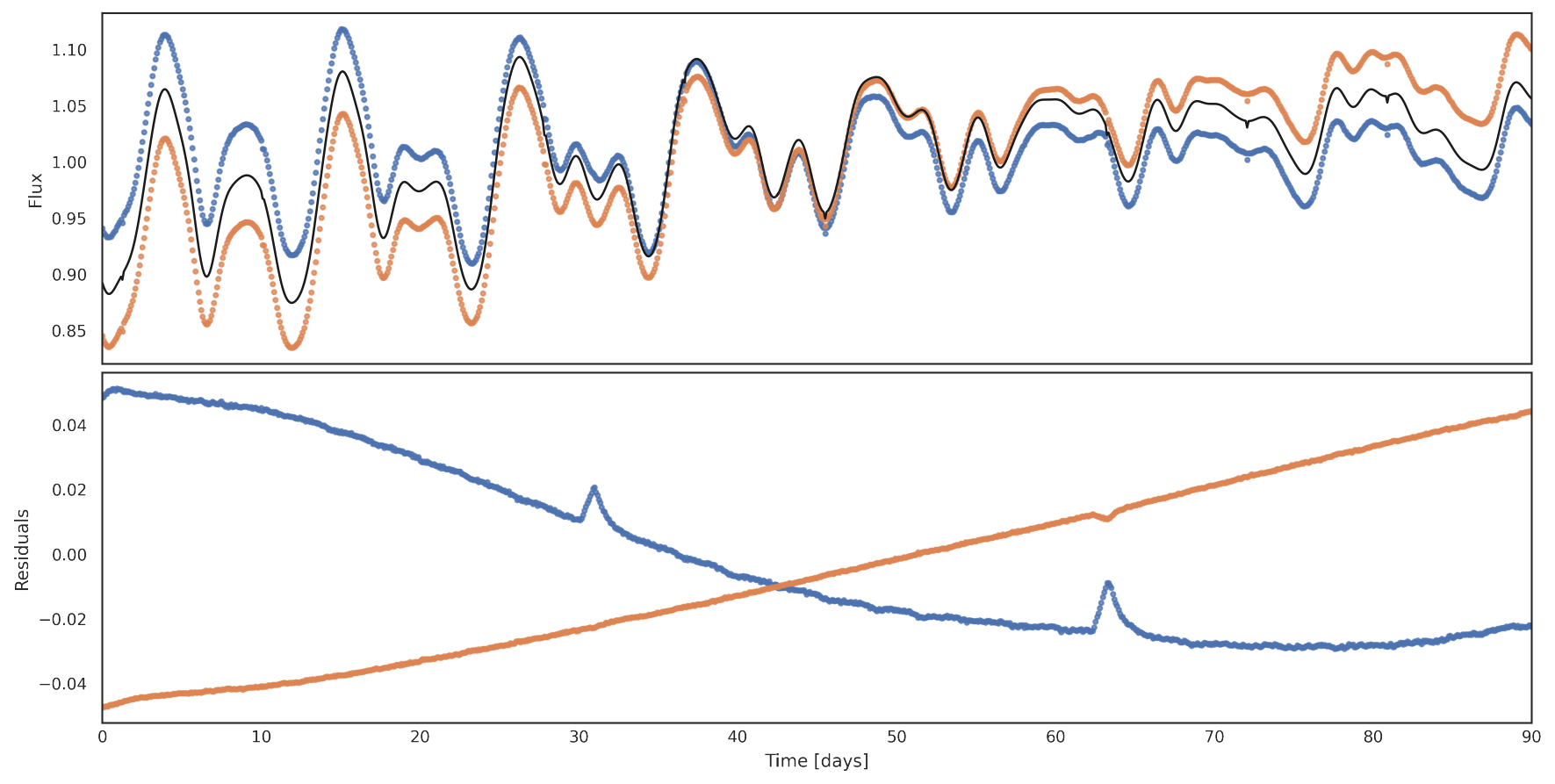}\\
    \includegraphics[width=0.98\textwidth]{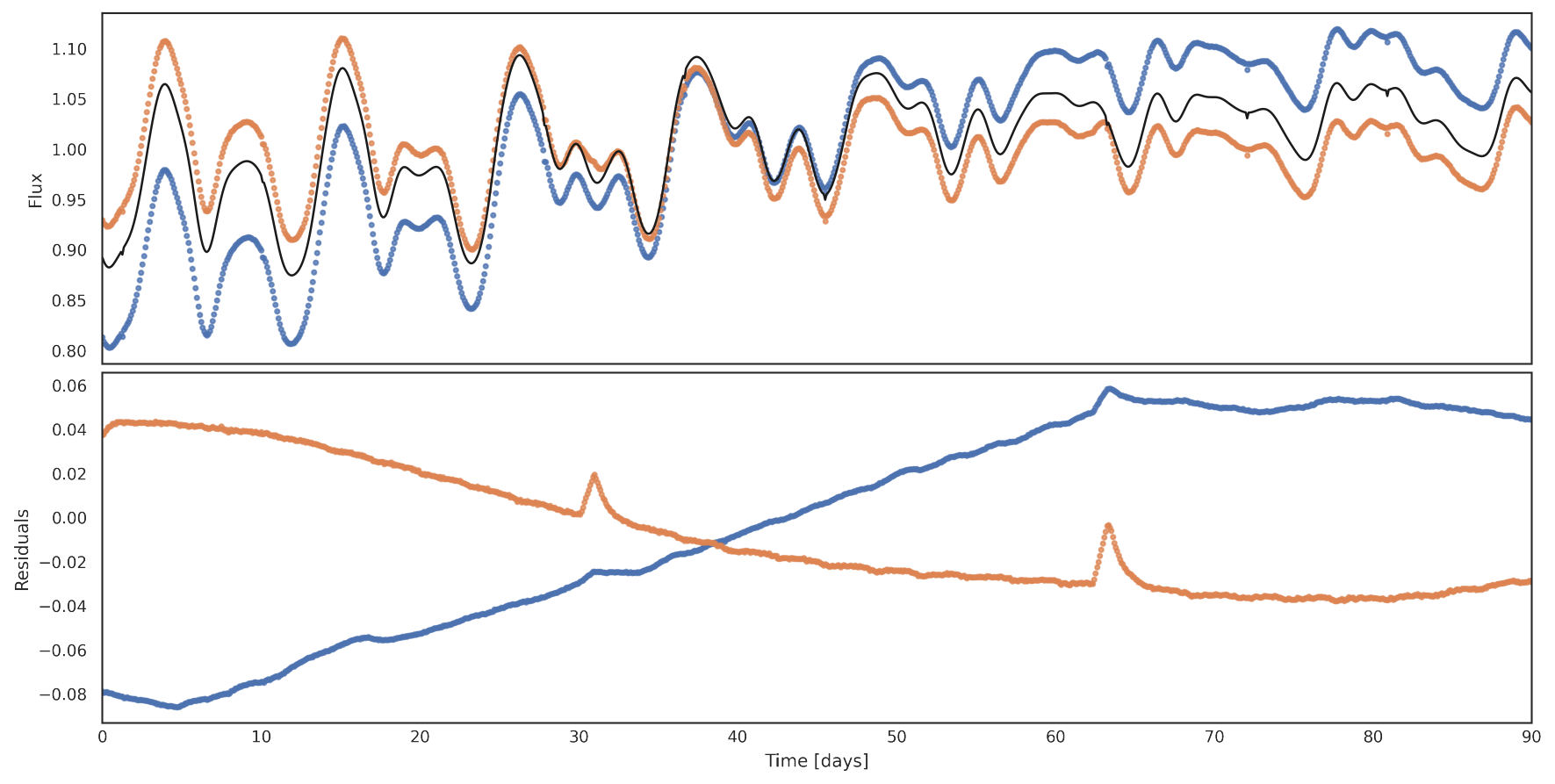}
   \caption{\oscar{
   \protect\republic\ (orange) and PDC-LS-like (blue) performance in the case of similar trends in the CBVs. 
   Top panel shows the result of perfectly known trends while bottom panel shows the results with the PCA-extracted trends.
   The black line shows the true astrophysical signal.
   We show the corresponding residuals for each case. 
   }
    }
    \label{fig:lccorrelated}
\end{figure*}

We then proceed to relax another key assumption, namely that the systematic trends are substantially different between one camera and the next. This could happen in a real-world scenario if the dominant sources of systematics affect multiple cameras at the same time, for example, due to changes in airmass or seeing (for ground-based observations from the same site), telescope motion (for multiple cameras on the same mount) or differential kinematic aberration (for multiple cameras sharing the same FOV).
\oscar{We note that from a theoretical point of view, repetition of the exact same trend across multiple cameras renders the least-squares problem ill-posed and this could cause the \republic\ algorithm to fail.}
To take this scenario to its extreme, we modified the procedure described in Sect.~\ref{sec:sim_trends} to simulate the trends, \oscar{forcing one of the four trends to have the same behaviour by all cameras. To do so we took one of the \kepler\ CBVs and we assign as the first trend injected to each light curve. The only variation is that for each of them we added a slightly white noise of 50\,ppm.
This scenario represents a stringent test case, where the trends are nearly identical, simulating conditions that are, in practice, less likely due to the natural variations expected in real observations where shared trends would exhibit more significant differences. 
The left panel of Figure~\ref{fig:trends_correlated} shows the trends used for this analysis for 6 of the simulated cameras. We can see that Trend 1 (blue line) has the same behaviour for all the cameras.
}
We then created another set of 1000 light curves using this new set of trends and applied the \republic\ correction as we did in Sect.~\ref{sec:idealcase}. 
\oscar{
The top panel of Figure~\ref{fig:lccorrelated} shows the \republic\ and PDC-LS corrections to one of the light curves.
We can see that both algorithms provide a light curve that resembles the real one. The general shape of the light curve is maintained, but both algorithms leave a residual trend. However, we note that \republic\ does better at correcting the bumps at $\sim 30$ and $\sim 60$ days}.

We then tested the behaviour of \republic\ on the same set of light curves, but using trends extracted from the light curves themselves using PCA as described in Sect.~\ref{sec:pcatrends}. The resulting trends are shown in the right panel
Figure~\ref{fig:trends_correlated}. As before, the PCA-recovered trends are qualitatively similar, but not identical, to the injected ones. Importantly, there are now small differences between the first trend (blue curve) in each camera, though the overall behaviour remains the same.
We applied the \republic\ algorithm to the light curves using these new PCA-extracted CBVs and found that in this case, \republic\ works nominally. The results, an example of which is shown in Figure~\ref{fig:lccorrelated}, are similar to the case where the trends are different across all cameras and the trends are extracted using PCA (Sect.~\ref{sec:pcatrends}). 
We note that the identical trends scenario does not affect the PDC-LS-like algorithm, as the PDC-LS correction is performed on a camera-by-camera basis. For completeness, we show in Figure~\ref{fig:lccorrelated} the PDC-LS-like correction in the case in which CBVs of different cameras have identical trends.

\section{Conclusions}

We have presented \republic, a new algorithm that can be used to correct instrumental systematics while preserving astrophysical signals in light curves from multi-camera surveys such as \plato. \republic\ exploits the fact that the astrophysical signal is the same across all cameras, but the systematics vary from camera to camera, to disentangle the two. Using simulated \kepler-like data, we showed that, in idealised conditions, when the systematic trends are known and weakly or uncorrelated, \republic\ outperforms a naive correction that ignores the intrinsic variability of the target star. 

In particular, \republic\ avoids the suppression of signals on timescales similar to the duration of the observations, which is a hallmark of standard systematic correction methods such as the widely-used PDC-MAP method. \oscartwo{Conversely, \republic\ is designed to preserve signals which are common to multiple cameras, so it will not remove systematics that have the exact same effect on multiple cameras, as those will be incoroporated in the "stellar" part of the model. Therefore, \republic\ and can be seen as a `conservative' alternative to the more `aggressive' PDC-MAP correction.}
For this reason, \republic\ is being implemented in the \plato\ L0-L1 pipeline (which will produce science-ready light curves) as an alternative to the baseline PDC-MAP algorithm. While the preservation of long-term astrophysical signals is not crucial for  applications such as transit detection or asteroseismology, it will be useful for the detection of rotation periods, particularly for slower rotators, as well as the study of longer-term phenomena such as active region evolution and activity cycles.

As with any systematics-correction method, the performance of \republic\ depends directly on the accuracy of the trends used to model the systematics. The tests presented in this paper either assumed that the trends were known a-priori, or were extracted from the light curves themselves using PCA. 
A possible extension of the \republic\ algorithm would be to solve for the systematic trends at the same time as for the astrophysical signal, over the entire ensemble of light curves at once. This may seem un-realistic, but the mathematical structure and scale of the resulting problem would not be dissimilar to that of the large linear models used in numerous machine learning applications.

The performance of \republic\ also depends critically on the degree of co-linearity between the systematic trends affecting the different cameras. If those are too similar, the algorithm becomes numerically unstable and the correction fails. In tests where only one of 4 trends was shared across all cameras, we found that using PCA to extract the trends from the light curves in each camera alleviates this problem significantly: it introduces subtle differences between the extracted trends in different cameras that break the degeneracy.

\section*{Acknowledgements}

O.B. and S.A. acknowledge support from the UK Science and Technology Facilities Council (STFC) under grants ST/S000488/1 and ST/R004846/1. 
This work made use of \texttt{numpy} \citep[][]{numpy}, \texttt{matplotlib} \citep[][]{matplotlib}, \texttt{seaborn} \citep{seaborn}, \texttt{pandas} \citep{pandas}, \texttt{sklearn} \citep[][]{scikit-learn}, \texttt{scipy} \citep{SciPy-NMeth} and \texttt{Pytransit} \citep{pytransit} libraries.
This work presents results from the European Space Agency (ESA) space mission
PLATO. The PLATO payload, the PLATO Ground Segment and PLATO data processing
are joint developments of ESA and the PLATO Mission Consortium (PMC). Funding for
the PMC is provided at national levels, in particular by countries participating in the
PLATO Multilateral Agreement (Austria, Belgium, Czech Republic, Denmark, France,
Germany, Italy, Netherlands, Portugal, Spain, Sweden, Switzerland, Norway, and United
Kingdom) and institutions from Brazil. Members of the PLATO Consortium can be found
at \url{https://platomission.com/}. The ESA PLATO mission website is
\url{https://www.cosmos.esa.int/plato}. We thank the teams working for PLATO for all their work.
O.B. acknowledges Dr. Baptiste Klein and Dr. Annabella Meech for their P supportive words that motivated me to finally complete this paper.

\section*{Data Availability}

The code used to perform the tests and figures in this manuscript is available in \url{https://github.com/oscaribv/REPUBLIC_paper_code}.



\bibliographystyle{rasti}
\bibliography{example} 








\bsp	
\label{lastpage}
\end{document}